# Lost-circulation diagnostics using derivative-based type-curves for non-Newtonian mud leakage into fractured formation

Rami Albattat[1], Marwa AlSinan[2], Hyung Kwak[2], Hussein Hoteit [1,*]

[1]King Abdullah University of Science and Technology (KAUST), Thuwal, Saudi Arabia

[2]Saudi Aramco, Dhahran, Saudi Arabia

## Abstract

Drilling is a requisite operation for many industries to reach a targeted subsurface zone. During operations, various issues and challenges are encountered, particularly drilling fluid loss. Loss of circulation is a common problem that often causes interruptions to the drilling process and a reduction in efficiency. Such incidents usually occur when the drilled wellbore encounters a high permeable formation such as faults or fractures, leading to total or partial leakage of the drilling fluids. In this work, a novel semi-analytical solution and mud type-curves (MTC) are proposed to offer a quick and accurate diagnostic model to assess the lost-circulation of Herschel-Bulkley fluids in fractured media. Based on the observed transient pressure and mud-loss trends, the model can estimate the effective fracture conductivity, the time-dependent cumulative mud-loss volume, and the leakage period. The behavior of lost-circulation into fractured formation can be quickly evaluated, at the drilling site, to perform useful diagnostics, such as the rate of fluid leakage, and the associated effective fracture hydraulic properties. Further, novel derivative-based mud-type-curves (DMTC) are developed to quantify the leakage of drilling fluid flow into fractures. The developed model is applied for non-Newtonian fluids exhibiting yield-power-law (YPL), including shear thickening and thinning, and Bingham plastic fluids. Proposing new dimensionless groups generates the dual type-curves, MTC and DMTC, which offer superior predictivity compared to traditional methods. Both type-curve sets are used in a dual trend matching, which significantly reduces the non-uniqueness issue that is typically encountered in type-curves. Usage of numerical simulations is implemented based on finite elements to verify the accuracy of the proposed solution. Data for lost circulation from several field cases are presented to demonstrate the applicability of the proposed method. The semi-analytical solver, combined with Monte Carlo simulations, is then applied to assess the sensitivity and uncertainty of various fluid and subsurface parameters, including the hydraulic property of the fracture and the probabilistic prediction of the rate of mud leakage into the formation. The proposed approach is based on a novel semi-analytical solution and type-curves to model the flow behavior of Herschel-Bulkley fluids into fractured reservoirs, which can serve as a quick diagnostic tool to evaluate lost-circulation in drilling operations.

## Introduction

Various industries are highly interested in drilling for different purposes, such as natural resources exploitation, $CO_2$ geological storage, nuclear waste, environmental remediation, and others [1]. To perform such operations, a variety of drill equipment is used, including drilling fluid that provides pressure control, cooling, lubrication, bit rotation, cuttings removal, among others. However, during drilling operations, many difficulties and challenges are encountered, sometimes causing environmental, economic, and safety issues. Unexpected problems are usually

Corresponding author: Hussein.hoteit@kaust.edu.sa

emerged due to the complexity of the nature of subsurface formations and wellbore components [2]. One of the most challenging problems is lost-circulation (LC) of drilling fluid. Loss of circulation is a pressing problem in drilling operations, including oil and gas and geothermal industries [3–8]. LC is typically subdivided into four categories in terms of the intensity of loss rate; seepage, moderate, severe, and total losses [9,10]. This challenging issue can disturb drilling operations and exacerbate non-productive time (NPT), which may lead to cost overrun depending on the loss type [11–17]. Wellbore circulation fluid can be lost into multiple mechanisms, assigned generally to four geological formation nature; porous matrix, large cavities, natural fractures, and drilling-induced fractures [18], as illustrated in **Figure 1**.

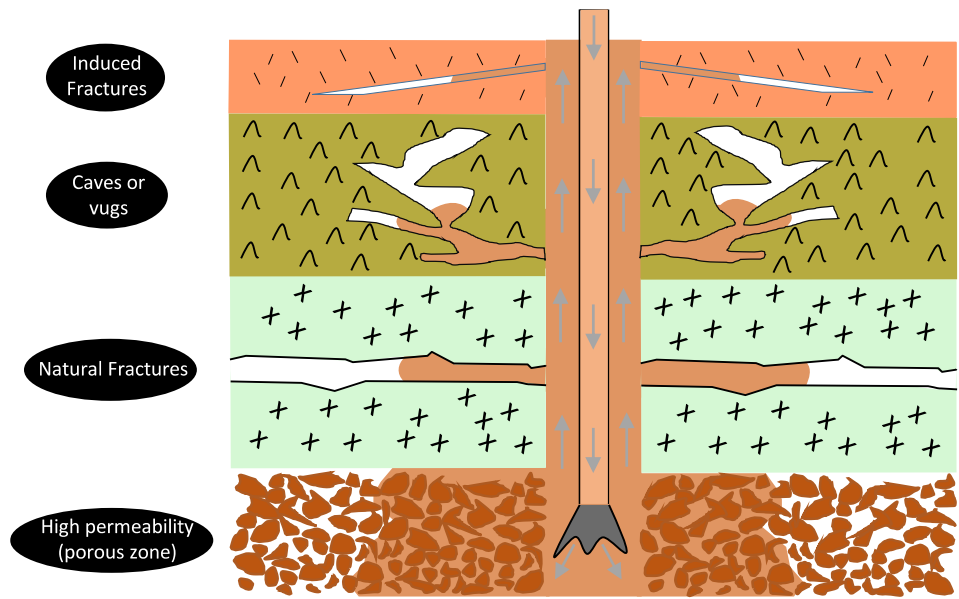

**Figure 1**: *Lost circulation may occur in rock formations with four different mechanisms. The arrows indicate the circulating-fluid direction from surface pumps then to drilling pipe, open-hole, wellbore annulus, and back to surface. The brown color represents the invasion mud from the wellbore to the surrounding formations.*

Lost circulation has been experienced in various places. For instance, it is reported that 35% of the total drilled wells encountered mud loss in Iran in carbonate rocks because of natural fractures [19], and in other places in the Middle East [20]. Carbonate formations are highly prone to mud loss due to presence of extensive natural fractures and LC in carbonates is much more challenging [21,22]. LC has also occurred in the Gulf of Mexico and other locations around the world [23–28]. Such drilling problems can significantly increase the NPT and operational cost, mainly in high-temperature regions [29–33]. Furthermore, mud loss may cause formation damage, safety hazards, and pollutes water tables [12–16,34,35]. Consequential problems due to LC are summarized in **Table 1** [10,36,37]. To develop a mitigation plan, general guidance is commonly used in LC. Seepage loss is often affordable to proceed drilling with no hindrance, whereas partial loss can often be crucial, and a decision is required either to cure the loss or to proceed. However, severe losses into thief zones and fractures often require immediate action to mitigate circulation loss. The focus of this paper is on modeling mud leakage into natural fractures, as illustrated in **Figure 2**, as it represents the most common severe scenario [38].

**Table 1**: *Problems induced due to lost-circulation from three drilling and completion jobs.*

| Cementing | Drilling | Completion/workover |
|---|---|---|

| Casing corrosion (external & internal) | Blowout and kill operations | Fluid loss |
|---|---|---|
| Bad zonal isolation | Downhole kicks | Non-productive time |
| Reduced safety | Environmental issues | Proximity formation damage |
| Reduced annular space coverage | Mud loss | Well loss |
| Demand for remedial cementing techniques | Time loss | More cost |
| Water table pollution | Stuck pipe | Wellbore integrity issues |
| | Poor cement job | |
| | Safety issues | |

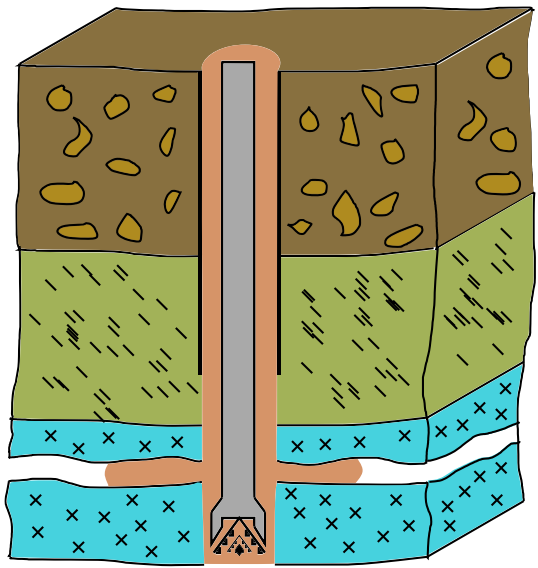

***Figure 2***: *Illustration of mud loss occurred as the drilling bit transverses a conductive natural fracture.*

Naturally fractured formation is often subjected to severe circulation loss [39–42]. To mitigate the loss, managed pressure drilling (MPD), redesign circulating fluid system, lost circulation material (LCM), and plugging thief zone are all available solutions. LCM is a standard effective solution practiced to mitigate LC [43–47]. For a given incident scenario, the selection of LCM, such as type, viscosity, and density, should be compatible with the hydraulic formation properties, geological formation type, and other factors [48,49].

Lost circulation often occurs suddenly and requires immediate action. Due to the time-scale constraint, a diagnostic tool that could be quickly applied on-site is needed. Such a tool should be based on accurate and efficient modeling methods to provide optimized solutions and what-if scenarios. In natural fracture, the fluid loss commences mainly with a sudden jump in flow rate followed by a slow and gradual decline in loss rate [50]. The ultimate total loss volume depends on many factors including, fracture extension, pore volume, fracture conductivity, and fluid mobilities [51,52].

One of the common approaches to predict loss decline curves is to use type-curves. This approach has been applied extensively in well-test [53], often used to evaluate flow dynamics and encountered formation properties serving both fluid flow loss and formation evaluation. Estimation of formation properties, determination of the best formation model, and identification of the transient flow patterns can be obtained from type-curves [54].

The development of analytical solutions regarding LC phenomenon and its formation evaluation has been attempted for years in the industry. For instance, Darcy's Law assumption was applied at steady-state conditions as an early attempt to model LC [55,56]. Semi-analytical transient flow assuming Newtonian fluid was then introduced to solve flow equation in natural fracture [57]. Later, an analytical solution was proposed by assuming a constant viscosity model for the mud [58]. Estimation of hydraulic fracture aperture size based on a particular Bingham plastic model was developed by [59]. A similar approach was adopted by other researchers [60,61]. Estimation of fracture aperture was achieved analytically after ignoring insignificant terms [62]. Extension to inclined fractures was proposed by [63]. An advanced solution of the flow equation in which the fluid follows Herschel-Bulkley fluid was introduced by multiple authors [42,64,65]. This approach, however, may produce inaccuracies for a particular problem because of some simplification in the derivation, and therefore, an improvement was proposed [66]. Nevertheless, a known problem with type-curves is related to the non-uniqueness of the solution. An effective approach to address this issue is to combine the type-curves with their corresponding derivative-based curves, similarly to those used in well testing [67–69]. The concept of derivative-based type curves has been used by [70] for Newtonian fluids in fractured formation with simplified fracture configurations. The inclusion of complex fractures often requires advanced numerical modeling [71–74].

In this work, we introduce new dimensionless groups and propose novel derivative-based mud type-curves (DMTC) for non-Newtonian fluids. The derivative-based curves are combined with the mud type-curves (MTC), reflecting the time-dependent dimensionless mud loss rate with its derivative. We discuss the mathematical derivation in detail for MTC and DMTC. High-resolution simulations with the finite element method were used to verify the semi-analytical solutions and the corresponding type-curves. We demonstrate the approach by analyzing several field cases. Finally, we perform a sensitivity analysis for the physical parameters to assess uncertainties.

**Problem Statement and Model**

The wellbore is assumed to intercept a horizontal infinite-acting fracture with an effective hydraulic aperture, $w$, as illustrated in **Figure 3**. Lost circulation occurs due to mud leakage from the wellbore into the fracture, driven by the pressure drop, $\Delta p$ between the well flowing pressure $p_f$, and the fracture pressure $p_{res}$. As the shear-thinning mud travels away from the wellbore, its apparent viscosity builds up as a function of the decreasing shear rate. Consequently, the mud mobility and the leakage rate reduce, leading eventually to a complete stall of the mud propagation [66,75].

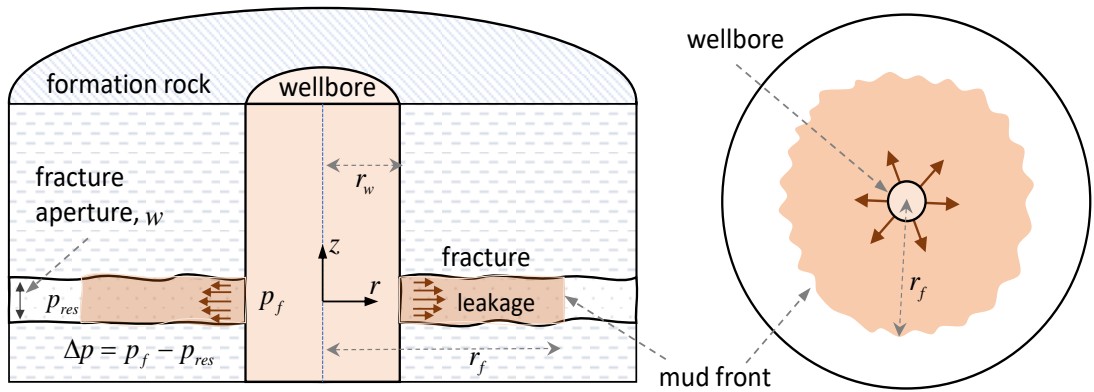

**Figure 3**: Illustration showing a side view (left) of a wellbore intercepting a fracture with an aperture $w$, and a top view (right) showing the mud invasion into the fracture, highlighted by the shaded radial area with an average radius $r_f$.

The radial shear stress component $\tau(z, r)$ of the Herschel-Bulkley fluid model is given by [76]:

$$\tau(z, r) = \tau_0 + m\left(\frac{dv_r}{dz}\right)^n,$$ (1)

Where $v_r$ is the radial velocity, $\frac{dv_r}{dz}$ describes the shear rate, $\tau_0$ is the fluid yield stress, $m$ is the consistency coefficient, and $n$ is the flow behavior index reflecting the rheological behavior of the mud, such that for $n < 1$, $n = 1$, and $n > 1$ the mud exhibits shear-thinning, Newtonian, and shear-thickening behavior, respectively.

The flow of the non-Newtonian mud within the fracture is modeled by the Cauchy equation of motion, that is,

$$\rho \frac{\partial \mathbf{v}}{\partial t} + \rho(\mathbf{v} \cdot \nabla)\mathbf{v} = \nabla \cdot (-p\mathbf{I} + \boldsymbol{\tau}) + \rho \mathbf{g},$$ (2)

Where $\rho$ is mud density, $\mathbf{v}$ is velocity vector, $t$ is time, $p$ is fluid pressure, $\mathbf{I}$ is the identity matrix, $\boldsymbol{\tau}$ is the shear stress tensor, and $\mathbf{g}$ is the gravitational acceleration.

For 1D radial flow, the pressure drop, $\Delta p$, is derived in terms of the total volumetric leakage rate $Q_{total}$ and the mud front distance $r_f(t)$, as follows (see **Appendix A**):

$$\Delta p = \frac{B\left(r_f(t) - r_w\right)}{2} + \frac{Q_{total}^n \left(r_f(t)^{1-n} - r_w^{1-n}\right)}{2(1-n)A} + \frac{1}{2} \int_{r_w}^{r_f(t)} \left(\sqrt{\left(B + \frac{Q_{total}^n}{r^n A}\right)^2 - 4D}\right) dr,$$ (3)

Where $A$, $B$, and $C$, are functions of the mud physical properties, $n$, $\tau_0$, and fracture aperture $w$.

### Analytical solution and type-curves

The derived equation from Eq. (3) for mud invasion front as a function of time becomes :

$$1 = \frac{B\left(r_f - r_w\right)}{2\Delta p} + \frac{Q_{total}^n \left(r_f^{1-n} - r_w^{1-n}\right)}{2\Delta p(1-n)A} + \frac{1}{2\Delta p} \int_{r_w}^{r_f(t)} \left(\sqrt{\left(B + \frac{Q_{total}^n}{r^n A}\right)^2 - 4D}\right) dr.$$ (4)

Eq. (4) can be rewritten in a compact form reflecting three dimensionless terms, such that:

$$\Psi\left(\mathbf{r_f}, r_w, \Delta p, n, w, \tau_0\right) + \Phi\left(\mathbf{r_f}, r_w, \Delta p, n, w, m, Q_{total}\right) + \Theta\left(\mathbf{r_f}, r_w, \Delta p, n, w, m, Q_{total}, \tau_0\right) = 1,$$ (5)

Where,

$$Term\ I:\ \Psi = \frac{B\left(r_f - r_w\right)}{2\Delta p}$$

$$Term\ II:\ \Phi = \frac{Q_{total}^n\left(r_f^{1-n} - r_w^{1-n}\right)}{2\Delta p\left(1-n\right)A}$$

$$Term\ III:\ \Theta = \frac{1}{2\Delta p}\int_{r_w}^{r_f(t)}\left(\sqrt{\left(B + \frac{Q_{total}^n}{r^n A}\right)^2 - 4D}\right)dr$$

(6)

In the above equations, the term $\Psi$ represents the stopping condition for mud invasion relative to fluid yield stress; $\Phi$ reflects the behavior of mud propagation before the fluid comes to a stop, while the third term $\Theta$ includes some effects of $\Psi$ and $\Phi$ but with less significance, as discussed below.

Eq. (5) shows a fractional relationship of the three dimensionless terms that sum up to unity. As one term decreases, the other terms increase so that their sum is always one. A volumetric illustration of $\Psi$, $\Phi$, and $\Theta$ and their typical behavior versus time is shown in **Figure 4**. The term $\Psi$ is proportional to the invasion radial distance $\left(r_f - r_w\right)$, and inversely proportional to $\Delta p$, which is essentially constant at a given depth. At later stages, as the mud propagates further away from the wellbore, $\Psi$ increases as a result of increasing $\left(r_f - r_w\right)$ (see Figure 4). On the other hand, the term $\Phi$ exhibits its highest value at the early time of mud loss. It then decreases with time, reflecting the behavior of the time-dependent leakage rate $Q_{total}^n$, which is initially high and then gradually declines with time. This mud-loss declining behavior is consistent with field observations [77]. As the mud loss gets to a stop, $\Phi$ vanishes, $\Psi$ peaks while $\Theta$ accounts for a minor balancing value.

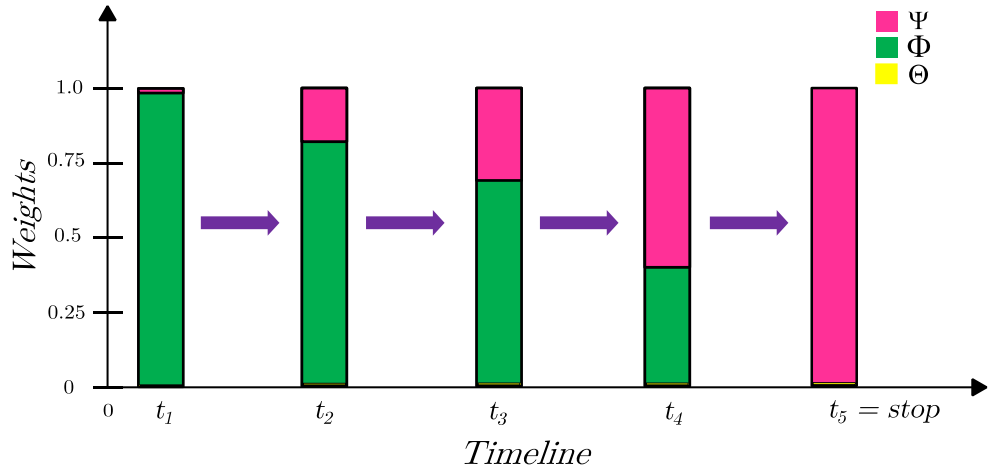

**Figure 4**: *Illustration showing a typical behavior of the three dimensionless terms* $\Psi$, $\Phi$, *and* $\Theta$ *as a function of time. The summation of the three terms is always one at all times.*

Since the contribution of the $\Theta$ term is insignificant in comparison with $\Psi$ and $\Phi$, we ignore it momentarily to analyze the governing dimensionless groups. Transforming Eq.(4) to a dimensionless form and after ignoring $\Theta$, one gets:

$$1 = \frac{1}{2}\left(\frac{2n+1}{n+1}\right)\left(\frac{\tau_0}{\Delta p}\right)\left(\frac{2r_w}{w}\right)(r_D - 1) + \frac{2^n}{1-n}\left(r_D^{1-n} - 1\right)\left(\left(\frac{2n+1}{n}\right)\left(\frac{m}{\Delta p}\right)^{\frac{1}{n}}\left(\frac{r_w}{w}\right)^{\frac{n+1}{n}} r_D \frac{dr_D}{dt}\right)^n,$$  (7)

Where the dimensionless mud-invasion radius $r_D$ and other dimensionless parameters are defined as:

$$r_D = r_f / r_w$$

$$t_D = t\beta$$

$$\alpha = \left(\frac{2n+1}{n+1}\right)\left(\frac{2r_w}{w}\right)\left(\frac{\tau_0}{\Delta p}\right)$$  (8)

$$\beta = \left(\frac{n}{2n+1}\right)\left(\frac{w}{r_w}\right)^{1+\frac{1}{n}}\left(\frac{\Delta p}{m}\right)^{\frac{1}{n}}$$

In the above equations, $t_D$ is dimensionless time, $\alpha$ is the stopping parameter, and $\beta$ is a flow parameter. Substituting Eq. (8) into (7), one obtains:

$$1 = \frac{\alpha}{2}(r_D - 1) + \frac{2^n}{1-n}\left(r_D^{1-n} - 1\right)\left(r_D \frac{dr_D}{dt_D}\right)^n.$$  (9)

With Eq.(8), the dimensionless mud-loss volume can be defined by,

$$V_D = \frac{V_m}{V_w} = \frac{\pi w r_f^2}{\pi w r_w^2} = \frac{r_f^2}{r_w^2} = r_D^2 - 1$$  (10)

Leading to,

$$r_D = \sqrt{V_D + 1}.$$  (11)

Taking the derivative with respect to $t_D$,

$$\frac{dr_D}{dt_D} = \frac{1}{2\sqrt{V_D + 1}} \frac{dV_D}{dt_D}.$$  (12)

Combining Eq. (11) and (12) into Eq. (9), we get,

$$1 = \frac{\alpha}{2}\left(\sqrt{V_D + 1} - 1\right) + \frac{2}{1-n}\left((V_D + 1)^{\frac{1-n}{2}} - 1\right)\left(\frac{dV_D}{dt_D}\right)^n.$$  (13)

Eq. (13) describes the dimensionless volumetric loss $V_D$ as a function of dimensionless time. Rewriting $\Psi$ and $\Phi$ in Eq. (13) in terms of the dimensionless groups,

$$\Psi = \frac{\alpha}{2}\left(\sqrt{V_D+1}-1\right)$$

$$\Phi = \frac{2}{1-n}\left(\left(V_D+1\right)^{\frac{1-n}{2}}-1\right)\left(\frac{dV_D}{dt_D}\right)^n \tag{14}$$

As $\Psi$ vanishes, solving $\Phi$ in Eq. (13) gives the mud loss volume with no fluid yield stress, as shown in **Figure 5**. This plot shows the solutions of Herschel-Bulkley fluid flow representing a dimensionless time-dependent mud-loss with assuming no stopping criterion. Therefore, dimensionless mud loss increases linearly with dimensionless time. The solid lines are selected based on varying flow behavior indexes $n$ starting from unity to zero. The yellow region is the spectrum solutions that all possible solutions exist in between.

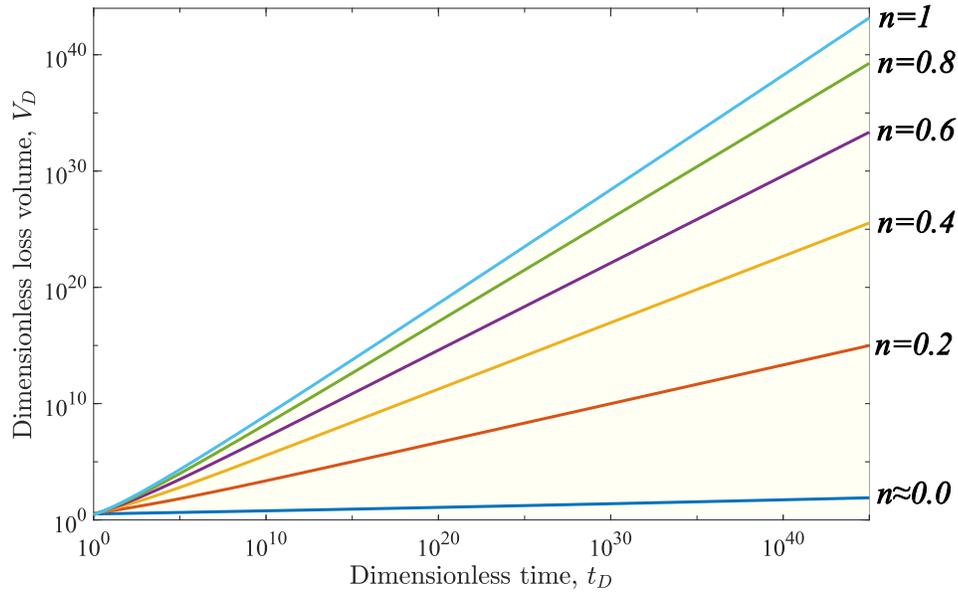

**Figure 5**: *Dimensionless mud-loss versus dimensionless time showing linear trend with the absence of yield-stress. The plots correspond to Bingham plastic case (n=1) and decreasing in the flow behavior index to almost zero, implying no flow. The yellow region represents the area of possible with variable n.*

Based on Figure 5, $V_D$ is almost linear in the log-log plots. Therefore, a power-law function is used to correlate $V_D$ to $t_D$, such that,

$$V_D = t_D^{c(n)} \tag{15}$$

The exponent $c(n)$, to be determined, stands for the slope observed in Figure 5 that is changing as a function of the flow behavior index $n$. The derivative of Eq. (15) with respect to $t_D$ yields,

$$\frac{dV_D}{dt_D} = c(n)t_D^{c(n)-1} \tag{16}$$

Substituting Eq. (15) into Eq. (16),

$$\frac{dV_D}{dt_D} = c(n)\frac{V_D}{t_D}$$

(17)

Placing flow behavior index $n$ as power on both sides,

$$\left(\frac{dV_D}{dt_D}\right)^n = \left(c(n)\frac{V_D}{t_D}\right)^n = c(n)^n\frac{V_D^{\,n}}{t_D^{\,n}}$$

(18)

Recalling Eq. (13) and ignoring $\Psi$ (Term I),

$$1 = \frac{2}{1-n}\left((V_D+1)^{\frac{1-n}{2}}-1\right)\left(\frac{dV_D}{dt_D}\right)^n$$

(19)

Substituting Eq. (18) into Eq. (19),

$$1 = \frac{2}{1-n}\left((V_D+1)^{\frac{1-n}{2}}-1\right)c(n)^n\frac{V_D^{\,n}}{t_D^{\,n}}$$

(20)

From Eq. (20), we can define,

$$\hat{V}_D = \frac{V_D^{\,n}}{1-n}\left((1+V_D)^{\frac{1-n}{2}}-1\right)$$

(21)

Thus, the combination of Eq.(21) into Eq. (20) yields,

$$1 = 2c(n)^n\frac{\hat{V}_D}{t_D^{\,n}}$$

(22)

Rearranging,

$$t_D^{\,n} = 2c(n)^n\,\hat{V}_D$$

(23)

In log-log scale, Eq. (23) becomes,

$$\log\left(t_D^{\,n}\right) = \log\left(\hat{V}_D\right) + \log\left(2c(n)^n\right)$$

(24)

Taking the derivative of Eq. (24) with respect to $\chi = \log\left(t_D^{\,n}\right)$,

$$\frac{d}{d\chi}\left(\log\left(t_D^{\,n}\right)\right) = \frac{d}{d\chi}\left(\log\left(\hat{V}_D\right)\right) + \frac{d}{d\chi}\left(\log\left(2c(n)^n\right)\right)$$

(25)

Since $c(n)$ is only a function of $n$, the last term in Eq. (25) vanishes. The first term on the left is equal to one, therefore, Eq. (25) simplifies to,

$$\frac{d}{d\chi}\left(\log\left(\hat{V}_D\right)\right) = 1$$

(26)

Substituting back the term $\chi$,

$$\frac{d\left(\log\left(\hat{V}_D\right)\right)}{d\left(\log\left(t_D^n\right)\right)} = 1 \tag{27}$$

Eq. (27) implies that plotting $\log\left(\hat{V}_D\right)$ versus $\log\left(t_D^n\right)$ exhibits a unit slope. Therefore, using $\hat{V}_D$ to define the type-curves is more convenient than the mud-loss volume $V_D$, as all curves at the early time can collapse to only one.

**Figure 6** shows examples of the type-curves, $\hat{V}_D$ versus $t_D^n$, for different values of $n$ and $\alpha$, representing different fluid types. The transformation into these dimensions is adequate in such a way that all cases of flow index must fall at the diagonal and then branch from this baseline depending on the stopping conditions. The stopping criteria follow a logarithmic stepwise, as shown in the **Figure 6**.

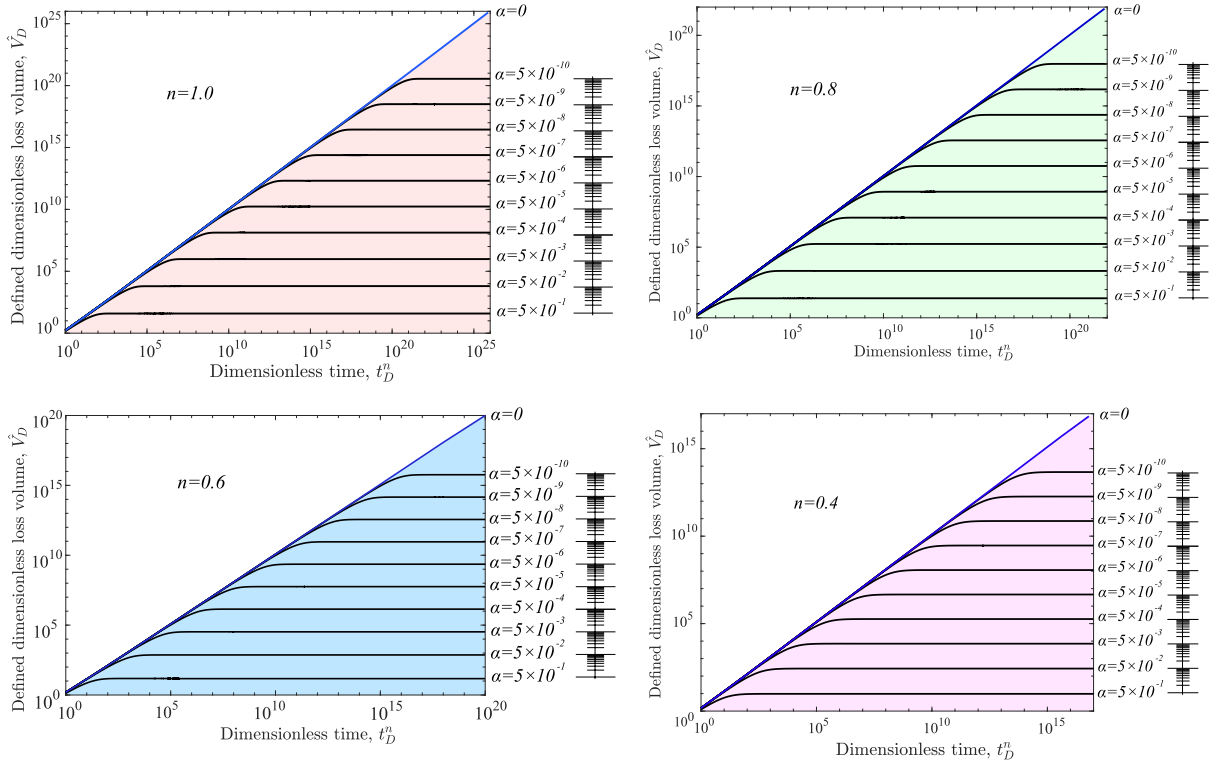

**Figure 6**: *Proposed type-curves provided by $\hat{V}_D$ versus $t_D^n$ in the log-log scale. The curves show almost a unit slope for different values of $n$ and $\alpha$. The diagonal lines represent the behavior where the flow does not exhibit a stopping criterion. The shaded region is the area where all the possible solutions can exist.*

The relationship between the dimensionless space to physical dimensional space is obtained from :

$$log\left(t_D^n\right) = nlog\left(t\beta\right) = nlog\left(t\right) + nlog\left(\left(\frac{n}{2n+1}\right)\left(\frac{\mathbf{w}}{r_w}\right)^{1+\frac{1}{n}}\left(\frac{\Delta p}{m}\right)^{\frac{1}{n}}\right)$$

$$= log\left(t^n\right) + nlog\left(\left(\frac{n}{2n+1}\right)\left(\frac{\Delta p}{m}\right)^{\frac{1}{n}}\right) + n\left(\frac{1+n}{n}\right)log\left(\frac{\mathbf{w}}{r_w}\right) \tag{28}$$

$$log\left(\hat{V}_D\right) = log\left(\frac{V_D^n}{1-n}\left(\left(1+V_D\right)^{\frac{1-n}{2}}-1\right)\right)$$

$$= log\left(\frac{1}{1-n}\right) + log\left(V_m^n\right) - log\left(\left(\pi r_w^2\right)^n\right) - log\left(\mathbf{w}^n\right) + log\left(\left(1+\frac{V_m}{\pi \mathbf{w} r_w^2}\right)^{\frac{1-n}{2}}-1\right) \tag{29}$$

Eqs. (28) and (29) provide the linking between the physical parameters to the type-curves. The fracture aperture $\mathbf{w}$ is an unknown that can be determined by matching observed data with the type-curves.

### Derivative-based solution

One general drawback about type-curves is the non-uniqueness of the solution. To address this issue, we develop additional derivative-based curves that can be applied in combination with the previously developed type-curves. The advantage of introducing the derivative-based curves is that they are more sensitive to capture deviations from the unit slope. For convenience, both sets of curves (type-curves and the derivative-based curves) should be of consistent scales to be plotted on the same axes.

The terms $\Psi$ and $\Phi$ (see Eq. (14)) vary between zero and one, as illustrated in **Figure 7**. $\Psi$ entails the stopping effect, whereas $\Phi$ starts at one due to the initial mud leak at a high rate. Afterward, the flow rate gradually reduces as the mud invasion comes to a stop, where $\alpha$ is the dimensionless parameter reflecting the stopping criterion.

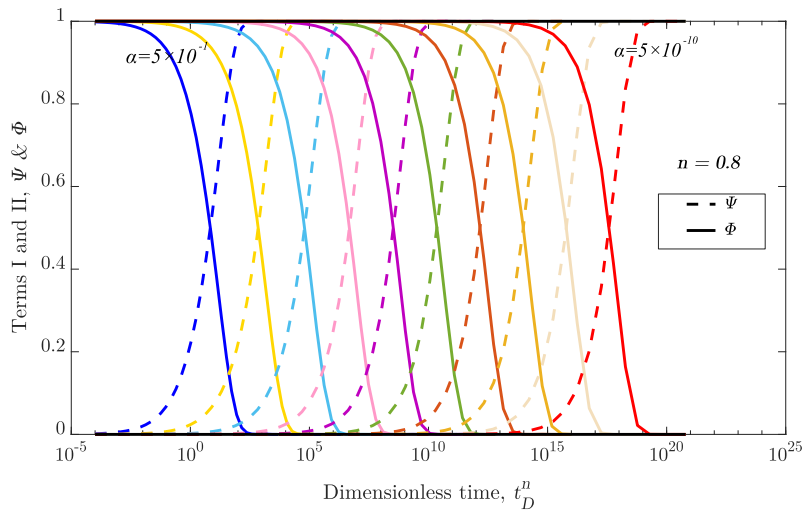

**Figure 7**: Typical behavior of $\Psi$ and $\Phi$ versus dimensionless time for different values of $\alpha$, as $\alpha$ decreases from left to right.

Since the term $\Phi$ incorporated the derivative $dV_D/dt_D$ (see Eq. (14)), we introduce the derivative-based parameter $\hat{V}_{DD}$, defined by:

$$\hat{V}_{DD} = \hat{V}_D \cdot \Phi \qquad (30)$$

Notice that $\hat{V}_{DD}$ is almost that same as $\hat{V}_D$ when $\Phi$ is close to one during the early time of mud leak. It then deviates from $\hat{V}_D$ towards zero, reflecting a change in the trend when $\Phi$ decreases as the mud invasion into the fracture come to a stop. **Figure 8** shows examples of the obtained mud type-cures (MTC) represented by $\hat{V}_D$, and the derivative-based mud type-curves (DMTC), represented by $\hat{V}_{DD}$, for different fluid types.

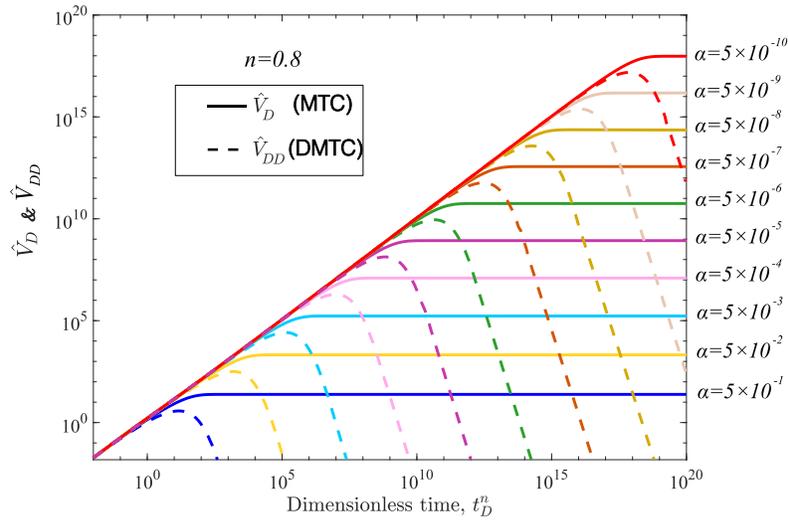

**Figure 8**: *Typical behavior of the proposed mud type-curves (MTC), represented by $\hat{V}_D$, and the corresponding derivative-based solution of type-curves (DMTC), represented by $\hat{V}_{DD}$. The solutions are for $n = 0.8$ with different values of $\alpha$.*

The linking equation between the derivative-based type-curves and the solution from physical space is given by:

$$
\begin{aligned}
log\left(\hat{V}_{DD}\right) &= log\left(\frac{\hat{V}_D}{1-n}\left(\left(1+V_D\right)^{\frac{1-n}{2}}-1\right)\left(V_D{}'\right)^n\right) \\
&= log\left(\frac{2}{(1-n)^2}\right) + log\left(V_m{}^n\right) - 2log\left(\left(\pi r_w{}^2\right)^n\right) - (3n+1)log\left(\mathbf{w}\right) + log\left(\left(1+\frac{V_m}{\pi \mathbf{w} r_w{}^2}\right)^{\frac{1-n}{2}}-1\right)^2 \\
&\quad + log\left(Q_{total}^n\right) - log\left(\left(\frac{n}{2n+1}\right)^n\left(\frac{1}{r_w}\right)^{n+1}\left(\frac{\Delta p}{m}\right)\right)
\end{aligned} \qquad (31)
$$

The right-hand side of Eq. (31) can be evaluated based on the mud-loss data and an effective fracture aperture $\mathbf{w}$, to be estimated by matching the observed data with the type-curves.

### Modeling procedure

The modeling workflow can be conducted in simple steps, which include developing the type-curves (i.e., the analytical solution) in Steps 1-3, preparing the field data in Steps 4-5, evaluating the match in Step 6, and repeating the steps until a satisfactory match is achieved. The procedure steps are given by :

1. For given fluid properties $\alpha$ and $n$ (if available), and an initial guess for $w$ (or any other uncertainty parameter), use Eq.(13) to calculate for dimensionless mud-loss volume $V_D$ as a function of dimensionless time $t_D$ .

2. Use Eq. (21) to plot $\hat{V}_D$ vs. $t_D^n$ in log-log scale, corresponding to MTC.

3. Use Eq.(30) and (14) to plot $\hat{V}_{DD}$ vs. $t_D^n$, which is the corresponding derivative-based mud type-curves (DMTC).

4. From the given mud-loss rates $Q_{total}$ , calculate $V_m$ vs. $t$ by using numerical methods.

5. Use Eq.(29) to plot the observed date in the form $\hat{V}_D$ vs. $t_D^n$ in log-log scale.

6. Evaluate the match between the type-curves and the processed field data. If the match is satisfactory, the procedure is complete.

7. If the match is not satisfactory, modify $w$ (and any other uncertainty parameters), and repeat from Step 1.

### Verification using simulations

To verify the proposed MTC and DMTC, a commercial simulator, COMSOL®, is used to solve the problem of non-Newtonian flow in a 2D radial system mimicking flow in a fracture. It should be noted that this simulator uses Navier-Stokes equations [78], whereas the representative governing equation is the Cauchy Equation of motion. To solve this discrepancy, we follow a technique proposed by Papanastasiou (1987), to the capture Cauchy equation by the Navier-Stokes equation through the Herschel-Bulkley fluid, such that,

$$\tau_{zr} = \tau_0 + m\left(\gamma\right)^n \tag{32}$$

Dividing all terms with the shear rate $\gamma$ ,

$$\frac{\tau_{zr}}{\gamma} = \frac{\tau_0}{\gamma} + m\left(\gamma\right)^{n-1} \tag{33}$$

Eq. (33) can be rewritten in the form,

$$\mu_{eff}\left(\gamma\right) = \mu_0 + m\left(\gamma\right)^{n-1} \tag{34}$$

The effective viscosity, $\mu_{eff}\left(\gamma\right) = \tau_{zr} \big/ \gamma$ , in Eq. (34) is a function of shear rate where $\mu_0 = \tau_0 \big/ \gamma$ is the viscosity generated from the fluid yield stress. To avoid the singularity in $\mu_0$, as $\gamma \to 0$, we introduced an exponential regularization function controlled by a parameter $m_p$,

$$\mu_{eff}\left(\gamma\right) = \mu_0\left(1 - e^{-m_p\gamma}\right) + m\left(\gamma\right)^{n-1} \tag{35}$$

The analytical solutions represented by the type-curves are compared with the results from solving the full Navier-Stokes equations corresponding to two test cases of non-Newtonian fluids with properties n=0.9 ( Figure 9 left) and

n=1 (Figure 9 right). The results show an excellent match between the numerical and analytical solutions. A detailed description of the simulation model and its validation is given in **Appendix B**.

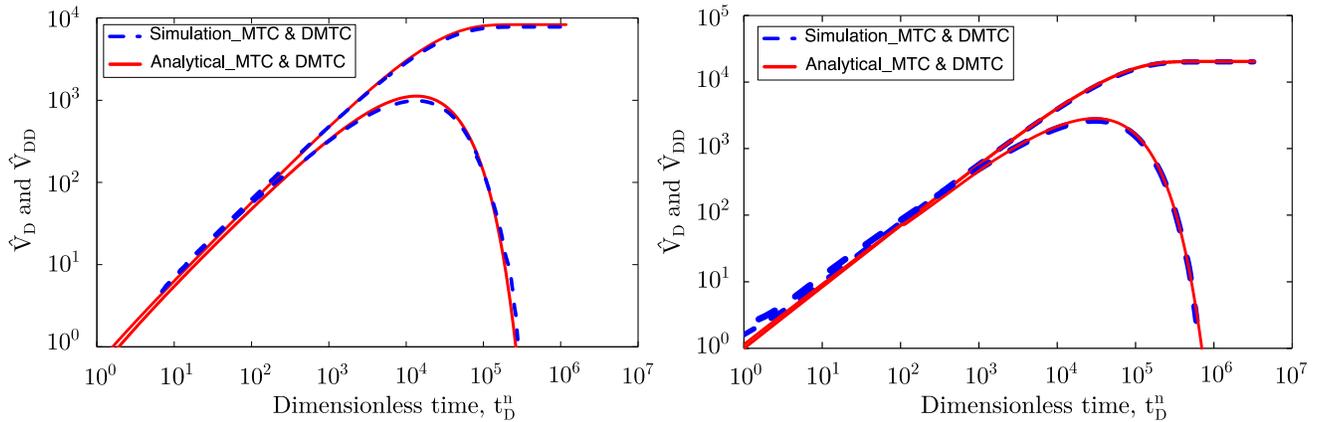

**Figure 9**: Comparisons of the proposed mud type-curves (MTC) and the derivative-based mud type-curves (DMTC) with the numerical results obtained by solving the full Navier-Stokes equations using COMSOL®, conducted for two test cases of fluid types corresponding to n =0.9 (left) and n=1 (right).

### Field Applications

To demonstrate the validity of the proposed model, we applied it to five wells that exhibited lost circulations in different fields [57,59,80,81]. The reported mud volume loss rates versus time for the five wells are shown in **Figure 10**. The relevant physical properties of the fluids are provided in **Table 2**. The well data, including the fluid rates, are converted to the dimensionless groups using Eqs. (28), (29), and (31). The processed data are then matched with the type-curves MTC, and DMTC. The matching curves provide an estimate of the effective fracture aperture and the ultimate mud loss volumes.

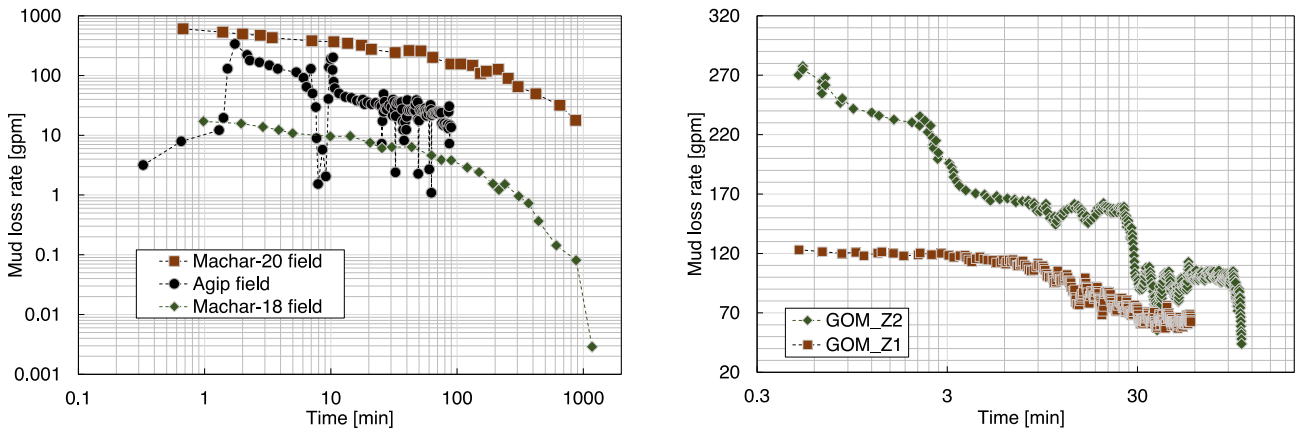

**Figure 10**: Volumetric flow rates for the five fields where the left figure shows data from two fields: Machar and Agip fields [57,59], while the right figure shows data for two wells from the Gulf of Mexico (GOM) [80,81].

**Table 2**: Well and field information with the fluid rheological properties used in the calculations.

| Property/Reference | Majidi et al. (2008) | | Liétard et al. (2002) | | Sanfillippo et al. (1997) |
|---|---|---|---|---|---|
| Area | Gulf of Mexico (GOM) | | Machar field, UK Central North Sea | | Agip fields |
| Incident name | Zone 1 | Zone 2 | Machar-20 | Machar-18 | Well A |
| Depth | 28,070 [ft] | 29,154 [ft] | 8468 [ft] | - | 11882.32 [ft] |
| Rock type | - | | Tor chalk | | limestones |
| Rheological parameters | $\tau_0 = 8.4$ [lb/100ft$^2$], $m = 0.08$ [lb/100ft$^2$.s$^n$], $n = 0.94$ | | $\tau_0 = 19.5$ [lb/100ft$^2$], $\mu_p = 30.5$ [cP], | $\tau_0 = 19.5$ [lb/100ft$^2$], $\mu_p = 48.1$ [cP], | $\tau_0 = 14.6$ [lb/100ft$^2$], $m = 0.042$ [lb/100ft$^2$.s$^n$], $n = 0.7$ |
| Wellbore radius | - | | $r_w = 8\ \frac{1}{2}$ [in] | | $r_w = 8\ \frac{1}{2}$ [in] |
| Pressure drop | 700-800 [psi] | | 1120 [psi] | 493.5 [psi] | 1010 [psi] |
| Rheology model | Herschel-Bulkley | | Bingham Plastic | | Herschel-Bulkley |
| Estimated average hydraulic aperture | $w = 0.69$ [mm] | $w = 0.84$ [mm] | $w = 0.616$ [mm] | $w = 0.425$ [mm] | $w = 0.33$ [mm] |
| Reported average hydraulic aperture | - | | - | - | 0.32-0.35 [mm] core sample |

The first field case corresponds to reported lost circulation incidents in Central North Sea, UK, at two wells: Machar-18 and Machar-20 [59]. The mud-loss rates and the mud physical properties corresponding to Bingham Plastic model, are provided in Figure 10a and Table 2, respectively. **Figure 11** shows the match for Machar-18 (left) and Machar-20 (right) with the analytical model represented by MTC and DMTC. The calculated fracture apertures corresponding to the match are 0.425mm and 0.616mm for Machar-18 and Machar-20, respectively. This aperture range is consistent with the one estimated by [59].

The data for the second field case is reported by [57], as shown in Figure 10a. The drilling mud properties, corresponding to a Herschel-Bulkley model, are given in Table 2. The best match of the analytical model with the data is shown in **Figure 12**, which corresponds to an effective fracture aperture of 0.33mm. This estimate is consistent with the aperture range of 0.32-0.35mm, suggested by Sanfillippo *et al.* based on experimental data of a core sample.

The third field case includes two wells in the Gulf of Mexico (GOM) [64,80]. The two wells correspond to two zones, named Z1 and Z2. Following a similar procedure, **Figure 13** shows the best match for the field data with the analytical model, which resulted in an estimate of the effective fracture aperture of 0.69mm and 0.82mm for well Z1 and Z2, respectively. Unlike the other test cases, the model, in this case, exhibits higher uncertainties in the match because of the lack of field data in establishing a clear deviation trend from the unit slope. This uncertainty implies the possibility of finding other solutions with equal matching quality. Therefore, uncertainty analysis is required, as discussed in the next section.

**Figure 11**: Match of the analytical solutions represented with MTC and DMTC with field data from the Central North Sea for well-18 (left) and well-20 (right). MTC is represented in solid blue lines, and the derivative solution (DMTC) is illustrated by dashed lines. The solid lines are the best fit, corresponding to 0.425mm and 0.616mm apertures, respectively. Well data are from [59].

**Figure 12**: Matching field data from Sanfillippo et al. (1997) with the analytical solution MTC, represented by the solid blue line, and the derivative-based solution DMTC, represented by the dashed line. The best fit corresponds to 0.33mm aperture.

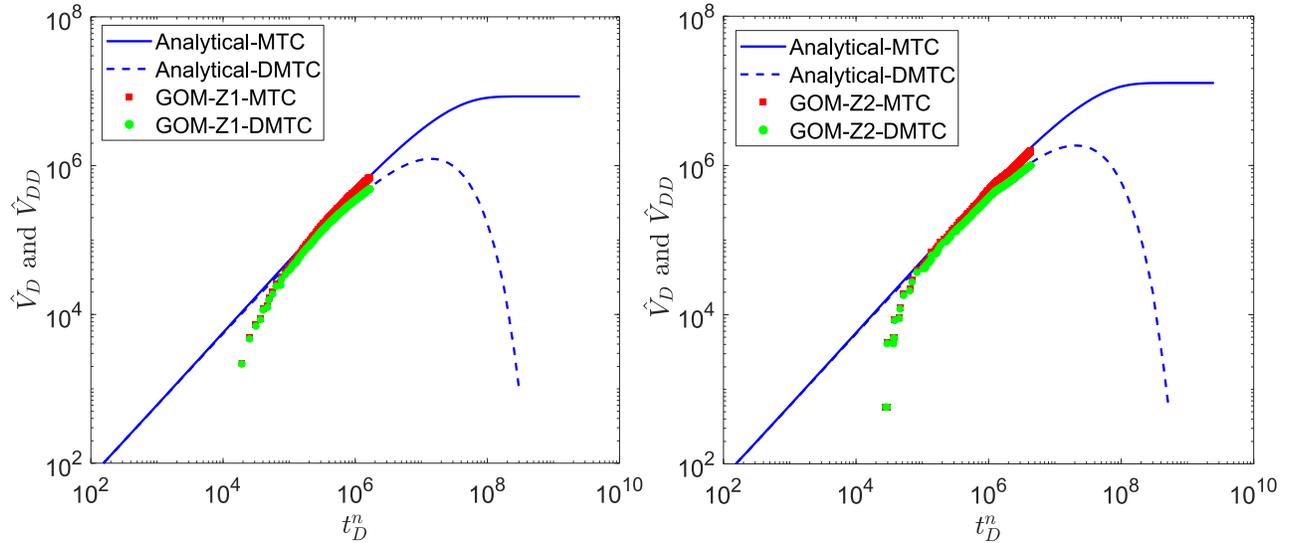

*Figure 13: Match of the analytical solutions represented by MTC and DMTC with field data from the Gulf of Mexico for well Z1(left) and well Z2 (right). MTC is represented in solid blue lines, and the derivative solution (DMTC) is illustrated by dashed lines. The solid lines are the best fit, corresponding to 0.69mm and 0.82mm apertures. Well data are from [64,80].*

### Uncertainty Analysis

An uncertainty-analysis procedure is combined with the analytical model to assess the probabilistic solutions and the corresponding predictions. The inclusion of the derivative-based solution (DTMC) provides an additional constraint to limit the range of possible solutions. For instance, **Figure 14a** provides the matching MTC solutions within 8% tolerance from the field data. The corresponding solutions ranges are [710.4-2853.3 bbl] for the ultimate loss volume and [0.652-0.719 mm] for the fracture aperture. By including DMTC and using the same tolerance (i.e., 8%) to select the matching MTC and DMTC curves, the corresponding range of solutions shrikes to [1148.5-2082.1 bbl] for mud loss, and [0.675-0.705 mm] for fracture aperture (see Figure 14**b**). These results, summarized in **Table 3**, which highlight the value of including DMTC to constrain the range of uncertainties.

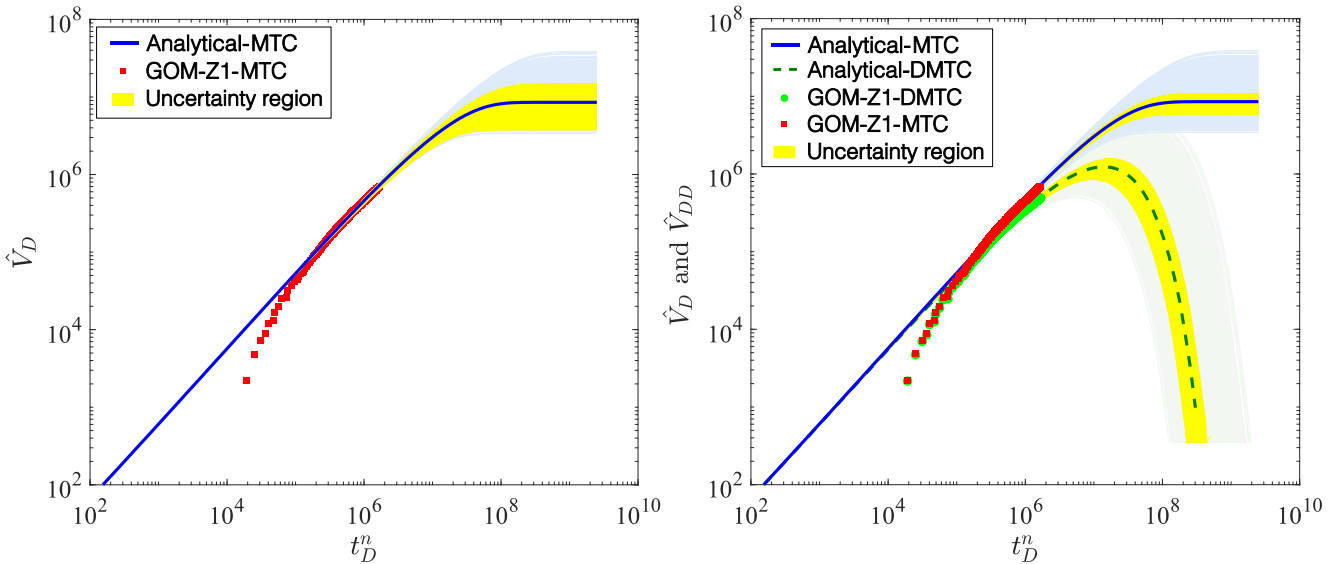

*Figure 14: Range of uncertainties in the solution by only using MTC to match the data within $L_2$ relative error of 8% (left), and the obtained range of uncertainties by including both MTC and DMTC to match the data within the same error, showing a more constrained solution and narrower uncertainty ranges.*

**Table 3**: *Uncertainty analysis reported for total mud-loss volumes and fracture apertures showing the effect of inclusion and exclusion of DMTC.*

| Property | Value | |
|---|---|---|
| Tolerance | 8% | |
| MTC uncertainty without DMTC | $V_m^{min} = 710.4$ [bbl] and $V_m^{max} = 2853.3$ [bbl] | |
| | $w^{min} = 0.652$ [mm] and $w^{max} = 0.719$ [mm] | |
| MTC uncertainty with DMTC | $V_m^{min} = 1148.5$ [bbl] and $V_m^{max} = 2082.1$ [bbl] | |
| | $w^{min} = 0.675$ [mm] and $w^{max} = 0.705$ [mm] | |

A sensitivity analysis study is further conducted to investigate the significance of various parameters on the model output [82–84]. Six rock and fluid parameters are considered in the study including, the pressure drop $\Delta p$, fracture aperture $w$, flow index $n$, consistency factor $m$, fluid yield stress $\tau_0$, and the wellbore radius $r_w$. Each parameter contributes to a certain degree to the mud loss amount. The influence of each parameter is evaluated by the normalized sensitivity matrix $\overline{\overline{s}}$, expressed as [85],

$$\overline{\overline{\mathbf{s}}} = \left(\nabla_{\mathbf{P}} \mathbf{E}\right) \cdot \frac{\mathbf{P}}{\mathbf{E}} ,\qquad(36)$$

Where $\mathbf{E}$ is the multivariable performance function with parameters $\mathbf{P}$, normalized by the term $\frac{\mathbf{P}}{\mathbf{E}}$. This normalized sensitivity function provides the relative impact of the parameters on the objective function [86,87]. The performance function is the time-dependent mud-front function $\mathbf{r}_f\left(t\right)$, which can be written as,

$$\mathbf{E} = \int_{t_\Omega} \mathbf{r}_f\left(t\right)\delta\left(t - t_p\right)dt = \mathbf{r}_f\left(t_p\right)\qquad(37)$$

Where; $\delta$ is Dirac delta function constrained on a selected time point $t_p$, corresponding to the mud front at tiem $t_p$. A average value is then used to capture the sensitivity $s_i$, that is,

$$s_i = \frac{1}{b_i - a_i}\int_{a_i}^{b_i} \overline{\overline{s}}\, dP_i ,\qquad(38)$$

Where the interval $\left[a_i, b_i\right]$ belongs to a region within a variable $P_i$.

**Figure 15** shows the variation of the calculated mud front as a function of several uncertainty parameters, corresponding to the fluid physical properties and fracture aperture, as shown in **Table 4**.

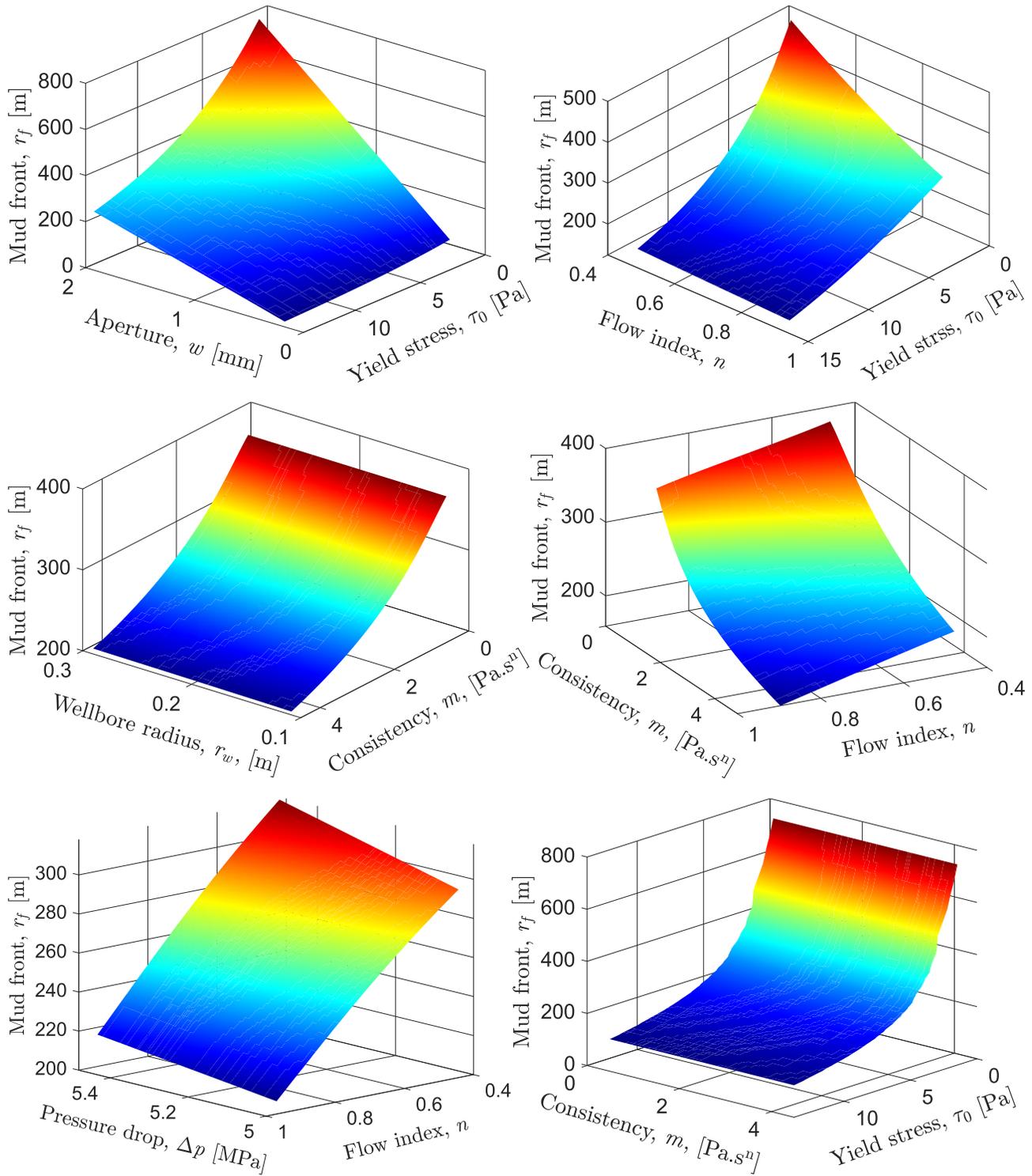

***Figure 15***: *Mud front behavior as a function of six uncertainty parameters plotted at* $t_p = 10^6 s$.

**Table 4**: *six parameters and their ranges used in the sensitivity study*

| Parameter | Range |
|---|---|
| Pressure drop, $\Delta p$ | 5-5.5 [MPa] |
| Hydraulic fracture aperture, $w$ | 0.1-2 [mm] |
| Fluid yield stress, $\tau_0$ | 1-15 [Pa] |
| Flow behavior index, $n$ | 0.4-0.95 |
| Consistency factor, $m$ | 0.1-5 [Pa.s$^n$] |
| Wellbore radius, $r_w$ | 0.1-0.3 [m] |

The calculated mud front demonstrates distinct behaviors as a function of the uncertainty parameters (**Table 4**). **Figure 16** illustrates a tornado plot showing the sensitivity assessment of the considered variables and their influence on the objective function, i.e., the mud front. As shown in **Figure 16**, the pressure drop and fracture aperture produced the highest effect on mud invasion, followed by fluid yield stress and flow index. The wellbore radius showed an insignificant effect, as expected. We should note that the sensitivity analysis is time-dependent, where the contribution of these parameters may change depending on the stage of mud loss.

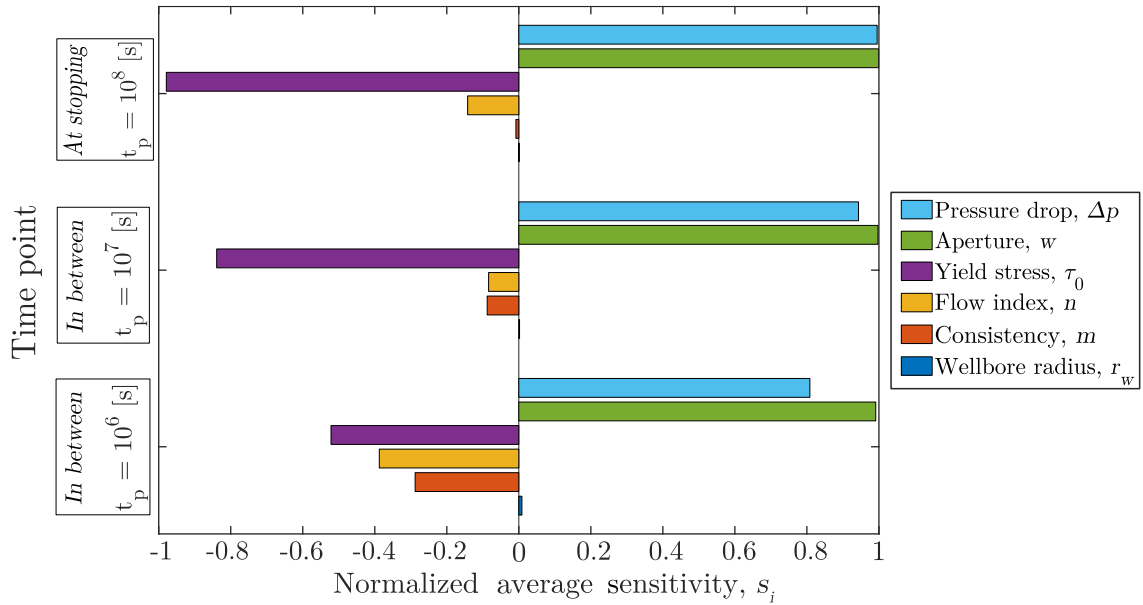

**Figure 16**: *Normalized average sensitivity analysis showing the relative contribution of various uncertainty parameters on the mud invasion to the fracture.*

## Conclusion

This work addresses the challenge of modeling lost circulation due to drilling mud leakage into natural fractures. Based on the Cauchy equation of motion for non-Newtonian fluids, we developed an analytical solution to model mud invasion into a natural fracture as a function of the pressure drop, fluid properties, and the effective hydraulic properties of the fracture. New mud type-curves (MTC) are then derived from the analytical solution. To address the issues of solution non-uniqueness, we proposed to combine MTC with derivative-based mud type-curves (DMTC).

A simple modeling procedure using MTC, DMTC, and the available field data is introduced. The analytical solution, including MTC and DMTC, are then verified using high-resolution finite-element simulations, which showed excellent agreement between the analytical and simulation results. Field cases, including mud-loss data from five wells, are then used to demonstrate the applicability of the proposed approach. Finally, a sensitivity analysis study is conducted to assess the most influential physical parameter on the objective function, corresponding to the mud invasion front. Fracture aperture and pressure drop showed the highest contributions, followed by the fluid yield stress and consistency factor. The objective of the sensitivity study is to demonstrate the applicability of the proposed solution in capturing relevant field uncertainties. The proposed analytical solution represents a novel modeling approach that could be efficiently implemented for field applications to provide diagnostics about mud loss and improve decision-making by providing a quick tool to perform predictions and what-if scenarios.

**Nomenclature**

| | | |
|---|---|---|
| $a_i$ | = | Upper bound of integral |
| $b_i$ | = | Lower bound of integral |
| $A$ | = | Defined constant |
| $B$ | = | Defined constant |
| $D$ | = | Defined constant |
| $\mathbf{E}$ | = | Performance function |
| $\mathbf{g}$ | = | Gravitational acceleration |
| $\mathbf{I}$ | = | Identity matrix |
| $\Delta p$ | = | Pressure drop |
| $\mathbf{P}$ | = | Multivariable parameters |
| $P_i$ | = | A variable parameter |
| $m$ | = | Consistency factor |
| $m_p$ | = | Non-physical regularization parameter |
| $n$ | = | Flow behavior index |
| $Q_{total}$ | = | Total volumetric flow rate |
| $r$ | = | Radial distance variable |
| $r_w$ | = | Hole inner radius |
| $r_D$ | = | Dimensionless radius |
| $\mathbf{r}_f$ | = | Mud invasion front radius variable |
| $dr$ | = | Derivative of radial variable |
| $s_i$ | = | Normalized average sensitivity at time point |
| $\bar{\bar{s}}$ | = | Normalized sensitivity matrix |
| $t$ | = | Time |
| $t_D$ | = | Dimensionless time |
| $t_p$ | = | Time selected point |
| $t_\Omega$ | = | Time domain |
| $\boldsymbol{v}_r$ | = | Velocity in radial direction |
| $V$ | = | Volume |
| $V_m$ | = | Mud loss volume |
| $V_m^{min}$ | = | Minimum mud loss volume |
| $V_m^{max}$ | = | Maximum mud loss volume |
| $V_D$ | = | Dimensionless mud loss volume |
| $\hat{V}_D$ | = | Defined dimensionless mud loss volume |
| $\hat{V}_{DD}$ | = | Dimensionless derivative solution of defined mud loss volume |
| $w$ | = | Fracture aperture |
| $w^{min}$ | = | Minimum hydraulic fracture aperture |
| $w^{max}$ | = | Maximum hydraulic fracture aperture |
| $x$ | = | x-axis |
| $y$ | = | y-axis |
| $z$ | = | z-direction |
| $\alpha$ | = | Defined dimensionless parameter for stopping condition |
| $\mu$ | = | Viscosity |
| $\mu_0$ | = | Viscosity from yield stress |
| $\mu_{eff}$ | = | Effective viscosity |
| $\rho$ | = | Fluid density |
| $\gamma$ | = | Shear rate |
| $\gamma_{min}$ | = | Minimum constrained shear rate |
| $\delta$ | = | Dirac delta function |
| $\tau_{zr}$ | = | Shear stress to z-direction on a surface plane normal to r-direction |
| $\tau_0$ | = | Fluid yield stress |
| $\Psi$ | = | Term I, stopping term condition |
| $\Phi$ | = | Term II, behavioral term flow |
| $\Theta$ | = | Term III, combination of Term I & II |
| $\chi$ | = | Logarithmic base 10 of dimensionless time to the power of flow behavior index |
| $\nabla_{\mathbf{P}}$ | = | Directional derivative for multivariable parameters |

**Abbreviation**

| | | |
|---|---|---|
| DMTC | : | Derivative-based Mud Type Curves |
| LC | : | Lost Circulation |
| LCM | : | Lost Circulation Materials |
| MTC | : | Mud Type Curves |
| MPD | : | Managed Pressure Drilling |
| NPT | : | Non Productive Time |
| GOM | : | Gulf of Mexico |
| Z1 | : | Zone 1 |
| Z2 | : | Zone 2 |

## Appendix A – Governing Equations

The general Cauchy equation used to describe the flow of non-Newtonian fluid is given by Eq.(2). We explicitly model the flow within the fracture geometry to avoid the limitation of the cubic law [88]. Taking into account the two main fundamental forces within a horizontal fracture, one gets,

$$\nabla \cdot \left( -p\mathbf{I} + \boldsymbol{\tau} \right) = 0 \tag{39}$$

Assuming a 1-dimensional (1D) radial system $r$, the radial shear stress component to z-direction becomes,

$$\tau(z, r) = z \frac{\partial p}{\partial r} \tag{40}$$

Herschel-Bulkley fluid model is defined,

$$\tau(z,r) = \tau_0 + m\left(\frac{dv_r}{dz}\right)^n \tag{41}$$

In the above equation, the shear rate symbolizes is $\frac{dv_r}{dz}$, reflecting the change of radial velocity $v_r$ in the z-direction.

Fluid properties are: fluid yield stress, $\tau_0$, consistency, $m$, and flow behavior index, $n$. Solving for the general velocity profile by combining Eq.(40) with (41) and applying boundary conditions for fracture aperture $z \in [0, w/2]$ leads to :

$$v_r(z) = \frac{n\left(-\frac{\partial p}{\partial r}\frac{w}{2} + \tau_0\right)\left(\frac{\frac{\partial p}{\partial r}\frac{w}{2} - \tau_0}{m}\right)^{1/n} + n\left(-\frac{\partial p}{\partial r}z + \tau_0\right)\left(\frac{\frac{\partial p}{\partial r}z - \tau_0}{m}\right)^{\frac{1}{n}}}{\frac{\partial p}{\partial r}(n+1)} \tag{42}$$

There are two velocity profile regions, plug and free. The plug region comprises the condition of no shear rate i.e., $\frac{dv_r}{dz} = 0$. Thus, after combining Eq. (40) and (41) with the plug region condition, one obtains,

$$z_{plug}\frac{\partial p}{\partial r} = \tau_0 \tag{43}$$

The velocity boundary conditions are,

$$v_r(z) = \begin{cases} v_{r,plug}(z), & for \ z \le z_{plug} \\ v_{r,free}(z), & for \ z_{plug} < z < \frac{w}{2} \\ 0, & for \ z = \frac{w}{2} \end{cases} \tag{44}$$

To get the free regions velocity $v_{r,free}$, Eq.(42) and (43) yield,

$$v_{r,free}(z) = \frac{n}{n+1}\left(z_{plug} - \frac{w}{2}\right)\left(\frac{\frac{\partial p}{\partial r}\left(\frac{w}{2} - z_{plug}\right)}{m}\right)^{1/n} + \frac{n}{n+1}\left(z - z_{plug}\right)\left(\frac{\frac{\partial p}{\partial r}\left(z - z_{plug}\right)}{m}\right)^{\frac{1}{n}} \tag{45}$$

Where $z_{plug}$ is fracture extension height of the plug region started from a centerline of the flow towards the upper or lower fracture walls. To get the plug region velocity $v_{r,plug}$, Eq. (43) and (45) demonstrate the plug velocity profile,

$$v_{r,plug}(z) = \frac{n}{n+1}\left(\frac{\tau_0}{\dfrac{\partial p}{\partial r}} - \frac{w}{2}\right)\left(\frac{\left(\dfrac{w}{2}\dfrac{\partial p}{\partial r} - \tau_0\right)}{m}\right)^{1/n} \tag{46}$$

The obtained two equations are for two velocity profiles, for plug $v_{r,plug}$ and free regions $v_{r,free}$, as follow:

$$v_{r,free}(z) = \frac{n}{n+1}\left(z_{plug} - \frac{w}{2}\right)\left(\frac{\dfrac{\partial p}{\partial r}\left(\dfrac{w}{2} - z_{plug}\right)}{m}\right)^{1/n} + \frac{n}{n+1}\left(z - z_{plug}\right)\left(\frac{\dfrac{\partial p}{\partial r}\left(z - z_{plug}\right)}{m}\right)^{\frac{1}{n}}$$

$$v_{r,plug}(z) = \frac{n}{n+1}\left(\frac{\tau_0}{\dfrac{\partial p}{\partial r}} - \frac{w}{2}\right)\left(\frac{\left(\dfrac{w}{2}\dfrac{\partial p}{\partial r} - \tau_0\right)}{m}\right)^{1/n} \tag{47}$$

The volumetric flow rates are defined by,

$$Q_{plug} = 4\pi r \int_0^{z_{plug}} v_{(r)plug}\, dz$$

$$Q_{free} = 4\pi r \int_{z_{plug}}^{w/2} v_{(r)free}\, dz \tag{48}$$

$$Q_{total} = Q_{plug} + Q_{free}$$

Computing total volumetric flow rate $Q_{total}$ by combining Eq.(48) and (47) leads to,

$$Q_{total} = \frac{4\pi r}{m^{1/n}}\left(\frac{dp}{dr}\right)^{1/n}\left(\frac{w}{2} - \frac{\tau_0}{\dfrac{dp}{dr}}\right)^{1/n+1}\left(\frac{n}{n+1}\frac{\tau_0}{\dfrac{dp}{dr}} + \frac{n}{2n+1}\left(\frac{w}{2} - \frac{\tau_0}{\dfrac{dp}{dr}}\right)\right) \tag{49}$$

After simplification,

$$0 = \left(\frac{dp}{dr}\right)^2 - \left(\frac{Q_{total}^n}{r^n\dfrac{(4\pi)^n}{m}\left(\dfrac{w}{2}\right)^{2n+1}\left(\dfrac{n}{2n+1}\right)^n} + \left(\frac{2n+1}{n+1}\right)\frac{\tau_0}{\dfrac{w}{2}}\right)\frac{dp}{dr} + \left(\frac{n-n^2}{n+1}\right)\left(\frac{\tau_0}{\dfrac{w}{2}}\right)^2 \tag{50}$$

Define the parameters,

$$A = \frac{(4\pi)^n}{m}\left(\frac{w}{2}\right)^{2n+1}\left(\frac{n}{2n+1}\right)^n$$

$$B = \left(\frac{2n+1}{n+1}\right)\frac{\tau_0}{w/2} \qquad (51)$$

$$D = \left(\frac{n-n^2}{n+1}\right)\left(\frac{\tau_0}{w/2}\right)^2$$

and solving for pressure profile distribution, the final form of mud front invasion as a function of time, $r_f(t)$ becomes,

$$\Delta p = \frac{B\left(r_f(t) - r_w\right)}{2} + \frac{Q_{total}^n\left(r_f(t)^{1-n} - r_w^{1-n}\right)}{2(1-n)A} + \frac{1}{2}\int_{r_w}^{r_f(t)}\left(\sqrt{\left(B + \frac{Q_{total}^n}{r^n A}\right)^2 - 4D}\right)dr \qquad (52)$$

Wellbore radius is $r_w$, $\Delta p$ is the wellbore/formation differential pressure. Further details about the mathematical derivation of Eq. (52) can be found in [66].

## Appendix B – Simulation Model

We use a commercial simulator (COMSOL®) to model the transport of non-Newtonian fluid governed by the Cauchy equation of motion given in given by Eq.(2). The viscous stress tensor is defined for incompressible fluid $\nabla \cdot \mathbf{v} = 0$ as,

$$\boldsymbol{\tau} = 2\mu\left(\nabla\mathbf{v} + \nabla\mathbf{v}^T\right) \qquad (53)$$

The shear rate is defined, $\boldsymbol{\gamma} = \frac{1}{2}\left(\nabla\mathbf{v} + \nabla\mathbf{v}^T\right)$. Combining Eq. (53) with Eq. (2), the Navier-Stokes equation is obtained,

$$\rho\frac{\partial\mathbf{v}}{\partial t} + \rho\left(\mathbf{v}\cdot\nabla\right)\mathbf{v} = \nabla\cdot\left(-p\mathbf{I} + 2\mu\left(\nabla\mathbf{v} + \nabla\mathbf{v}^T\right)\right) + \rho\mathbf{g} \qquad (54)$$

The Herschel-Bulkley fluid model is given by Eq. (33) and (34). The effective viscosity, $\mu_{eff}(\gamma) = \frac{\tau_{zr}}{\gamma}$, in Eq. (34) is a function of shear rate where $\mu_0 = \frac{\tau_0}{\gamma}$ is the viscosity generated from the fluid yield stress. To avoid the singularity in $\mu_0$, as $\gamma \to 0$, we introduced an exponential regularization function controlled by a parameter $m_p$, such that,

$$\mu_{eff}(\gamma) = \mu_0\left(1 - e^{-m_p\gamma}\right) + m(\gamma)^{n-1}$$

The viscosity equation is used as proposed by Papanastasiou (1987). The governing equation is linked with the fluid equation by the effective viscosity, i.e., $\mu = \mu_{eff}(\gamma)$. To illustrate the behavior of the regularization (see **Figure B-1**), we plot the shear stress versus shear rate for different values of $m_p$, where the regularized function converges to the actual Hersche-Bulkley model as $m_p$ or the shear rate increases. This step is critical to avoid numerical instability in the simulation model.

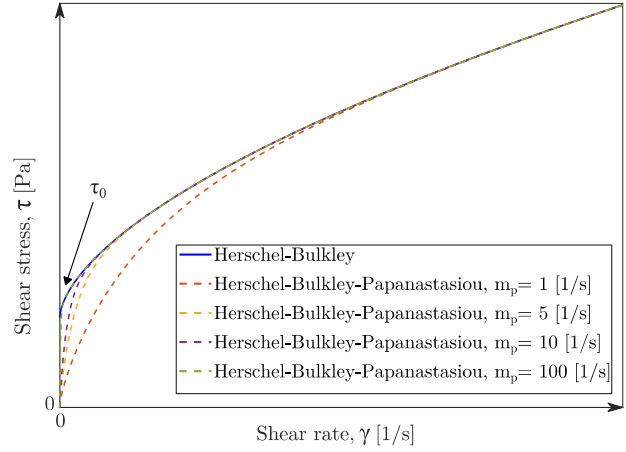

***Figure B-1***: *Behavior of the modified shear stress as a function of the regularized parameter $m_p$, used to avoid numerical instability in the simulator.*

The magnitude of shear rate is given by,

$$\gamma = \sqrt{2\gamma : \gamma} \tag{55}$$

Define the contraction operator "**:**" as,

$$\mathbf{a} : \mathbf{b} = \sum_i \sum_k a_{ik} b_{ik} \tag{56}$$

To avoid division by zero, we use a minimum shear rate $\gamma_{\min}$ as,

$$\gamma = |\gamma| = \max\left(\sqrt{2\gamma : \gamma}, \gamma_{\min}\right) \tag{57}$$

*Boundary conditions*

We use no-slip boundary conditions and no leakage through the fracture walls demonstrated as follow,

$$\mathbf{v} \cdot \mathbf{n} = 0$$
$$\left[-p\mathbf{I} + \boldsymbol{\mu}\left(\nabla\mathbf{v} + \nabla\mathbf{v}^T\right)\right]\mathbf{n} = 0 \tag{58}$$

For the prescribed pressure inlet and outlet, the simulator uses stress conditions as,

$$-p + 2\mu\frac{\partial v_n}{\partial n} = F_n \tag{59}$$

Where $\frac{\partial v_n}{\partial n}$ is the derivative of the normal velocity component. Here, $-p = F_n$ is used. For low Reynolds numbers, the tangential stress conditions becomes,

$$\mu\frac{\partial v_t}{\partial n} = 0 \tag{60}$$

Where $\frac{\partial v_t}{\partial n}$ is the normal derivative of the tangential velocity component.

**Figure B-2** shows the viscosity profiles at two different times after mud invasion starts in the fracture. As the mud travels away from the wellbore (see **Figure B-3**), viscosity increasing as a result of reduced shear thinning. The pressure profiles are shown in **Figure B-4**. The mud loss rate versus time is presented in **Figure B-5,** which illustrates how the mud comes to a stop. Further details of the simulation model and its validation with experimental data are found in [65].

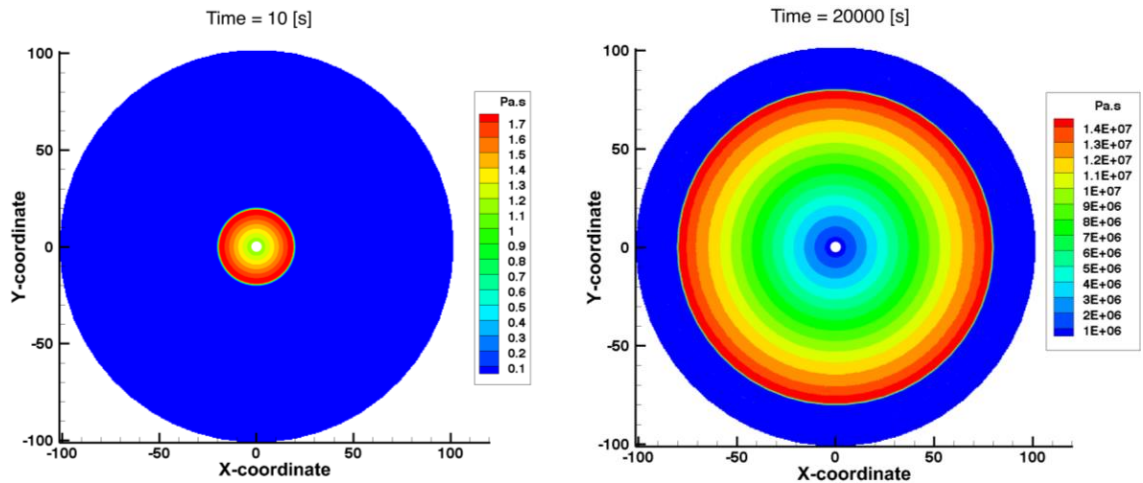

**Figure B-2**: *Viscosity maps at different times showing how the fluid viscosity increases as the mud travels away from the wellbore.*

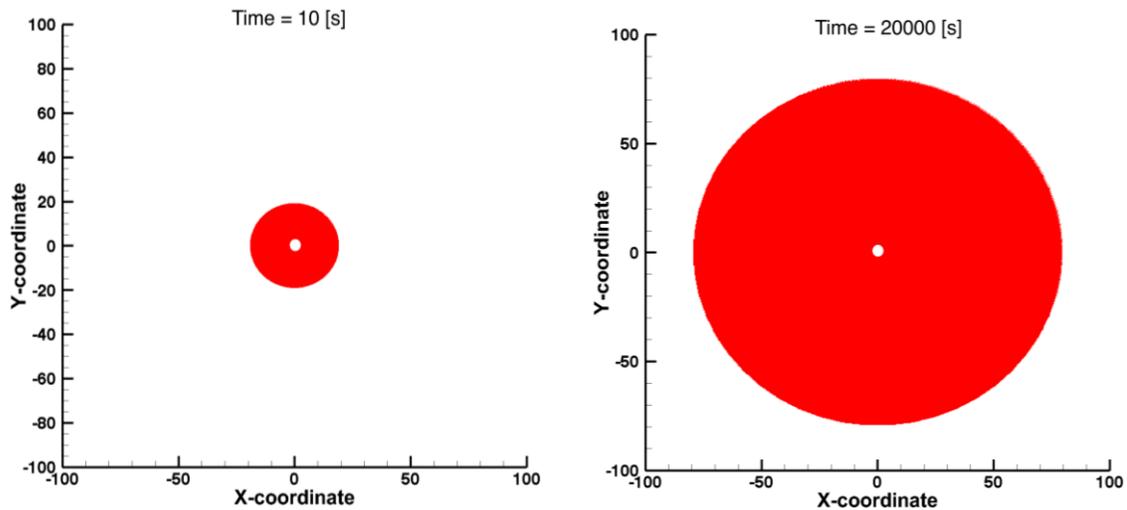

**Figure B-3**: *Mud front locations at two different times.*

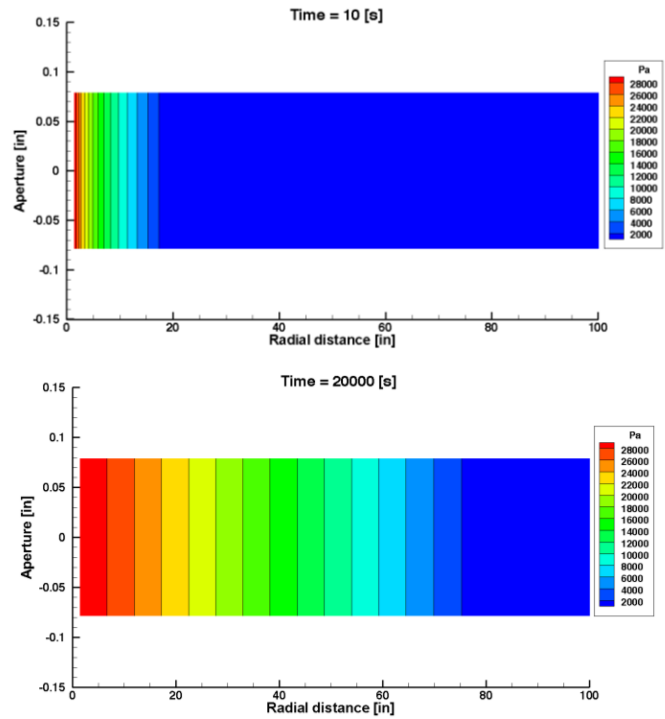

***Figure B-4****: Vertical cross-sections, showing the pressure profiles in the fractures at different times.*

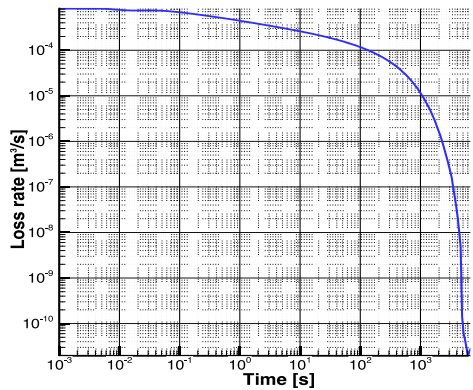

***Figure B-5****: Mud loss rate versus time, showing the rate decreasing to zero as the mud front comes to a stop.*